# Splitting of photo-luminescent emission from nitrogen-vacancy centers in diamond induced by ion-damage-induced stress

P. Olivero[1,2,3], F. Bosia[1,2,3], B. A. Fairchild[4], B. C. Gibson[4], A. D. Greentree[4,5], P. Spizzirri[4,6] and S. Prawer[4]

E-mail: olivero@to.infn.it

[1] Physics Department and "Nanostructured Interfaces and Surfaces" Centre of Excellence, University of Torino, via P. Giuria 1, 10125 Torino, Italy
[2] Consorzio Nazionale Interuniversitario per le Scienze Fisiche della Materia (CNISM), Italy
[3] Istituto Nazionale di Fisica Nucleare (INFN), sezione di Torino, Italy
[4] School of Physics and Melbourne Materials Institute, University of Melbourne, Victoria 3010, Australia
[5] Applied Physics, School of Applied Sciences, RMIT University, Victoria 3001, Australia
[6] The Melbourne Centre for Nanofabrication, Victoria 3168, Australia

**Abstract.** We report a systematic investigation on the spectral splitting of negatively-charged, nitrogen-vacancy ($NV^-$) photo-luminescent emission in single crystal diamond induced by strain engineering. The stress fields arise from MeV ion-induced conversion of diamond to amorphous and graphitic material in regions proximal to the centers of interest. In low-nitrogen sectors of a HPHT diamond, clearly distinguishable spectral components in the $NV^-$ emission develop over a range of ~4.8 THz corresponding to distinct alignment of sub-ensembles which were mapped with micron spatial resolution. This method provides opportunities for the creation and selection of aligned $NV^-$ centers for ensemble quantum information protocols.

## 1. Introduction

Diamond is rapidly becoming one of the most exciting solid-state platforms for quantum information processing (QIP) [1, 2]. This is due to a combination of factors, although one of the most important is the comparatively simple optical coupling to fluorescent colour centers with large optical dipoles and, in the case of the negatively-charged nitrogen-vacancy (NV⁻) center, a long-lived ground-state coherence [3].

Although much of the interest in NV⁻ based QIP is centered on room-temperature processes and isolated centers, there is also considerable interest in applying ensemble QIP protocols to the optical transition of inhomogeneously broadened NV⁻ centers. Indeed the dipole moment and transition frequency for NV⁻ are comparable to that of rubidium, meaning that translation of protocols developed for vapour cells is natural. Protocols using ensembles were initially considered for the first NV⁻ quantum computing [4, 5] and optical quantum non-demolition experiments [6]. There have been observations of long-lived ground state coherence [7], coherent population trapping [8], magnetometry [9] and magnetic coupling between NV⁻ ensembles and superconducting circuits [10, 11].

Despite the interest in ensemble processing with NV⁻, there are major challenges to be overcome, which place NV⁻ at a disadvantage when compared with atomic vapors. One of the largest problems is the inhomogeneous linewidth and the nature of this inhomogeneity. The inhomogenous linewidth of the optical transitions in NV⁻ typically varies with strain [12], implantation strategy [13, 14] and electric field [15] leaving the job of controlling the linewidth as a major goal of NV⁻ engineering. Coupled to this problem is that at room temperature, the linewidth appears dominated by spectral diffusion, although exceptional emitters with almost lifetime-limited emission can be discovered [16-20], and there are no reports of ensembles of purely homogenously broadened NV⁻ centres.

Here we show a new approach to spectrally separating aligned subsembles of NV⁻. By engineering permanent strain fields into the diamond lattice through ion-implantation, we demonstrate that it is possible to resolve the orientationally inequivalent NV⁻ subsembles based upon their spectral properties. These results reveal new opportunities for the use of aligned, inhomogeneously broadened ensembles of NV⁻ centers for quantum information processing that is distinct from more traditional methods of applying external uniaxial strain to a lattice.

In the study of defect-related optical transitions in crystals, the application of uniaxial stress is a well-established technique to investigate the irreducible representations of both the vibronic [21] and electronic states involved in the transitions and the point group of the optical centers [22, 23]. Since the 1960s, experiments were conducted to study the effect of uniaxial mechanical stress along the <100>, <110> and <111> directions on the spectral properties of the zero-phonon-line (ZPL) absorption and photoluminescence (PL) emission of characteristic defect-related centers in diamond, such as for example the neutral vacancy GR1 transition at $\lambda = 741$ nm [24-28], the neutral nitrogen-vacancy ($NV^0$) transition at $\lambda = 575$ nm [29, 30], the H3 line at $\lambda = 503$ nm [31, 32] and the H4 vibronic band at $\lambda = 496$ nm [33, 34]. In particular, absorption measurements of the ZPL $NV^-$ transition under uniaxial stress were performed allowing the attribution of this band to a transition between the $A_1$ (ground) and E (excited) electronic states of a trigonal center [12] and more recently to explore the 1046 nm transition [35].

With a thorough understanding of the effect of strain on the splitting and broadening of the ZPL emission, the opportunity now arises to use these effects to monitor internal stresses in artificial diamonds. In flame-grown CVD polycrystalline diamond films, the splitting of the ZPLs of the 533 nm and the $NV^0$ emissions were directly measured and mapped across the cubo-octaedral crystallites using cathodoluminescence (CL) [36]. The 533 nm emission was studied in greater detail, while the $NV^-$ emission was not acquired since the center is not CL-active [37]. The effect of stress on PL emissions from $NV^0$ ($\lambda = 575$ nm) and Si ($\lambda = 738$ nm) was also correlated with Electron Spin Resonance (ESR) measurements from the substitutional nitrogen center (P1) in both undoped and nitrogen-doped CVD diamond films [38]. Similarly, the linewidth of the GR1 neutral-vacancy-related emission at 741 nm was employed to qualitatively monitor the strain in a range of brown-colored type IIa diamonds before and after thermal annealing [39]. Stress-related ZPL shifts of the $NV^0$ emission were also recently measured in diamond nanocrystals using cathodoluminescence micro-mapping [40].

Here, we report the direct observation of the splitting of $NV^-$ PL emission engineered by the mechanical stress induced by MeV ion implantation. The implantation of energetic ions into diamond induces damage and amorphization, with associated swelling [41]. Here we show that the intense stress fields generated by such distortions induce a splitting in the $NV^-$ PL emission that can be mapped with micrometric spatial resolution in regions extending beyond the implanted regions. In the present experiment, PL mapping was performed at cryogenic temperatures (T = 80 K) on type Ib diamonds which had been implanted with 2 MeV $He^+$ ions, both before (i.e. as-implanted) and after thermal annealing. This allowed the direct observation of stress-induced splitting of the $NV^-$ emission at variable distances from the implanted region, which was also correlated with both finite element method (FEM) simulations of the mechanical stress and Raman experimental data.

## 2. Experimental

The sample under examination was a $3.5{\times}3.5{\times}1.5$ mm$^3$ type Ib ($10$ ppm $\leq$ [N] $\leq 100$ ppm) synthetic high-pressure-high-temperature (HPHT) single crystal diamond (Sumitomo). The sample was cut along the <100> crystal direction and optically polished on the two opposite large faces. Artificial diamonds produced with this technique are structured in macro-sectors that develop during the HPHT growth from a single seed; depending on their orientation, they are characterized by different impurity concentrations (N, Ni, etc.) and consequently they display distinct luminescence features [42]. As shown in the cathodoluminescence map of figure 1, the sectors are relatively large and structured along the crystallographic directions. In comparison with the "main" sector (A), the "strip" sectors extending near the sample edges (B) are characterized by lower nitrogen concentration, while the sectors near the sample corners (C) contain high concentrations of impurities (N, Ni, Fe…), as confirmed by photoluminescence (PL) mapping [43, 44].

The sample was implanted with 2 MeV He$^+$ ions at the MP2 microbeam line of the 5 U NEC Pelletron accelerator at the University of Melbourne. A region of $280{\times}125$ μm$^2$ extending across the "main" and "strip" sectors was irradiated in the region highlighted by the red square in figure 1, as shown in figure 2. An implantation fluence of $1{\times}10^{17}$ cm$^{-2}$ was uniformly delivered by raster scanning the ion microbeam; the ion current was $\sim$2 nA and the sample was implanted at room temperature. Using the SRIM 2008.04 Monte Carlo simulation code [45], we estimated the linear damage profile, expressed as the number of vacancies formed per incoming ion at a given depth. The simulation was performed by setting the atomic displacement energy to 50 eV [46] and, as shown in figure 3, the structural damage is mainly concentrated at the end of ion range in the material ($\sim$3.5 μm). A crude estimation of the damage density in the heavily damaged layer can be obtained by multiplying the damage linear density (expressed in vacancies ion$^{-1}$ cm$^{-1}$) by the implantation fluence (expressed in ions cm$^{-2}$), thus yielding the volumetric vacancy density. This linear approximation is only suitable for low damage densities as it significantly over-estimates the real values at high damage densities where it does not account for non-linear mechanisms such as self-annealing, ballistic annealing and defect interaction [47, 41]. Accordingly, we have employed a more realistic description of the damage process as described below. Nonetheless, it is worth noting that the linear approximation gives a density of $\sim$2$\times$10$^{23}$ vacancies cm$^{-3}$ at the ion end of range, indicating that the sample has been implanted at a damage level that significantly exceeds the amorphization threshold in this region [48-52, 41].

The ion beam is characterized by a Gaussian point spread function. Therefore, to achieve a sharper edge between the irradiated and unirradiated areas, the implanted area was masked on its right side with the cleaved edge of a monocrystalline Si wafer. As shown in the inset of figure 2, the masked edge exhibits a sharper transition from the implanted to the unimplanted area although we might expect that there will also be scattering beneath the implant edge [53]. By taking into account the spatial resolution of the imaging technique, we estimated the sharpness of the edge as being better than ~1 μm, which corresponds to the spatial resolution of the PL mapping techniques employed to characterize the sample (see below). As reported in figure 2, we will adopt a spatial reference convention so that the $x$ and $y$ axes lay on the sample surface and correspond respectively to the directions perpendicular and parallel to the edge of the implanted region while the $z$ axis extends perpendicularly to the sample surface along the depth direction. The blue arrow in figure 2 highlights the direction of the PL scan to locate the "strip" sector reported (reported below, see figure 8).

After ion implantation, the sample was thermally annealed with the purpose of forming a well-defined graphitic layer that would exert a stress field in the surrounding regions. The substrate was annealed for 1 hour at a temperature of 800 ºC, which is believed to be optimal for the formation of NV$^-$ aggregates [37], in a forming gas (4% H$_2$ in Argon) ambient; particular care was taken to control the annealing atmosphere to avoid oxygen contamination which would have resulted in surface etching of the diamond structure.

Photoluminescence maps at micrometer resolution were performed on a Renishaw RM1000 micro-Raman spectrometer on regions surrounding the implanted area. The $\lambda = 514.5$ nm emission of an Argon laser was employed as the excitation source after having been focused on the sample surface to a micrometer-sized spot by means of standard microscope optics with around 0.5-1 mW incident on the sample. The collection optics were configured for Raman backscattering measurements and comprised a notch filter for incident scattered light removal. The system was also equipped with computer controlled precision xy stages to translate the sample under examination for two-dimensional mapping with ~0.5 μm spatial resolution.

The spectrometer was equipped with a long-working-distance 50× objective lens (0.7 NA) and a stage mounted cryostat allowing the acquisition micro-PL spectra and maps at liquid nitrogen temperature (T = 80 K). The use of a single grating with either 1200 or 1800 lines mm$^{-1}$ coupled with a CCD array allowed the fast acquisition of maps with a spectral resolution that was adequate for photo-luminescence measurements over a broad spectral range (500-800 nm). For both gratings, the spectral resolution of the system was evaluated as better than ~0.17 THz (i.e. ~0.2 nm) at a wavelength of 551 nm, by measuring the full width at half maximum of the first-order Raman transition from same diamond sample (i.e. ~6 cm$^{-1}$), thus fully adequate to measure the spectral features of interest (see below).

Using this configuration and the standard aperture (50 μm) of the confocal acquisition system, the probed depth inside the sample was estimated to be ~1-2 μm. This implies that in the following analysis, the acquired signal should be considered as the average of the distribution of optical centers in the depth direction.

High resolution micro-Raman mapping measurements were performed on a Dilor XY microspectrometer. The basic instrumental configurations (excitation wavelength, co-axial focusing/collection system, sample scanning system) are similar to that described for the PL mapping with the following notable differences: the sample objective lens was of a high magnification (80×), measurements were performed at room temperature and spectra were acquired in a triple-additive grating mode allowing for a higher spectral resolution over a narrower spectral range. The spectral width of the first-order Raman line in a high quality diamond crystal was measured as 1.8 cm$^{-1}$. The laser power incident on the sample after focusing was ~1.5 mW while confocal parameters determined a probe depth which was estimated to be within the ~2-3 μm range.

Profilometry measurements were performed on the implanted sample both before and after thermal annealing to determine the swelling at each stage of sample processing. Profiles were obtained using an Ambios XP stylus profiler at a speed of 0.01 mm s$^{-1}$, force = 0.05 mg and filter setting 2 for each run the system which was calibrated using a 186.2 nm vertical standard. A typical surface roughness of the sample before implantation was ~5 nm.

## 3. Numerical simulations

To evaluate the stress dependence of the ZPL emission splitting and broadening of luminescent centers, numerical Finite Element Method (FEM) simulations were performed to assess the stress fields established in diamond upon ion implantation. The stress derives from a decrease in mass density and mechanical stiffness occurring in the material as a consequence of the ion damage to the crystal structure and resulting volume expansion of the amorphized regions. This volume variation is proportional to the mass density variation and is partially inhibited by the mechanical reaction of the surrounding undamaged material. Due to the non-trivial geometry of the problem, FEM simulations are best suited to adequately model this constrained volume expansion.

The mass-density change in the amorphized region is not uniform, since it arises from the damage depth profile in all three dimensions (i.e. the depth profile in figure 3 convolved with the implantation aperture). To correctly estimate the mass density (and hence volume) change, a recently introduced phenomenological model was adopted to describe the damage process in the crystal lattice at high implantation fluences [47]. The model is inspired by the work reported in [54] and takes into account the concentration of ion-induced vacancies with a simple linear approximation for the probability for a newly created vacancy to recombine with a self-interstitial:

$$P_{rec}(F,z) = \frac{\rho_V(F,z)}{\alpha} \, ,$$

(1)

where $P_{rec}$ is the recombination probability at a given depth $z$ and implantation fluence $F$, $\rho_V$ is the vacancy density in the material at depth $z$. $\alpha$ is an empirical parameter accounting for the defect recombination probability, i.e. it represents the saturation vacancy density in correspondence of which the recombination probability approaches 1 and therefore no further vacancies are induced in the structure. By solving the associated differential equation, we obtain an exponential saturation of the vacancy density for high implantation fluences. This relationship can be extended to the density variation in the damaged material as follows:

$$\rho(z) = \rho_d - (\rho_d - \rho_{aC}) \left[ 1 - e^{-\frac{\lambda(z)F}{\alpha}} \right] \, ,$$

(2)

where $\rho_d = 3.52$ g cm$^{-3}$ and $\rho_{aC} = 2.1$ g cm$^{-3}$ [41] are respectively the mass densities of pristine diamond and fully-amorphized carbon and $\lambda(z)$ is the linear density of induced vacancies per incoming ion as derived from SRIM. Starting from SRIM simulations for the 2 MeV He implantation, the depth profile of the mass density variation in the as-implanted sample can therefore be derived (as shown in figure 3). Consistent with the approach adopted with the modeling of the mass-density variation, we assume that the mechanical properties of the as-implanted diamond (i.e. Young's modulus $E(z)$ and Poisson's ratio $\nu(z)$) also vary between the corresponding values of diamond ($E_d = 1220$ GPa, $\nu = 0.20$) and amorphous carbon ($E_{aC} = 21.38$ GPa, $\nu = 0.45$) with the same trend reported in (2) for the vacancy density. Following this approach, we obtained the three-dimensional distributions of mass density and mechanical properties in the as-implanted sample, which we subsequently used as input in the FEM model to estimate the deformation and stress fields established in the sample [47].

To model the stress field after thermal annealing, for simplicity we invoke a simple binary approximation for the material composition. For regions where the vacancy density is below the graphitization threshold ($D_C$) we assume the lattice recovers to a crystalline structure. Conversely when the vacancy density exceeds $D_C$ we assume that the material is fully converted to graphite. Hence, we modeled the depth profiles of the material properties (density, mechanical parameters) as step-like functions varying between values relevant to diamond and graphite ($\rho_g = 2.1$ g cm$^{-3}$, $E_g = 10$ GPa, $\nu_g = 0.31$), as shown in figure 3, where the dimensions of the graphitic region are set by the extent of the critically damaged region. With the annealed sample, the obtained three-dimensional mass-density distribution and spatially varying mechanical properties are input into the FEM numerical simulations, to estimate the deformation and stress fields established in the sample [55].

We note that our models include only two free empirical parameters, that are important at different stages of the processing: before annealing, $\alpha$ empirically accounts for the probability of an ion induced vacancy/interstitial pair recombining with existing defects in the material, while for the post-annealing case $D_C$ is the critical density of vacancies above which the damaged structure converts to graphite upon thermal annealing. Note that whilst in principle these numbers should be known constants, in practice there is sufficient ambiguity in the literature as to their values, and hence treating them as parameters is appropriate. The other parameters, such as the mass density and the mechanical properties of diamond, amorphized carbon and graphite, were taken from the accepted literature values [47, 55, 41].

The three-dimensional FEM model was adopted to calculate all stress components in the unimplanted regions close to the edge of the implanted area that result from the constrained expansion of the nearby buried damaged layer. A free mesh with tetrahedral elements was adopted and refined in the region of interest (i.e. the edge of the implanted region). In both cases (i.e. before and after thermal annealing) the non-uniform constrained expansion of the implanted crystal was directly determined by the local variations of mass-density and mechanical parameters and fitting applied to $\alpha$ and $D_C$.

## 4. Results and discussion

We first report on the damage-induced swelling of the diamond. As mentioned above, MeV ion damage induces a significant distortion of the crystal structure with accompanying swelling and induced pressure in the material [56, 54, 57-60]. Figure 4 shows profilometry scans of the implanted sample before and after annealing, taken across the masked edge. Thermal annealing has the effect of increasing the swelling since the buried layer has been damaged beyond $D_C$, and therefore the expansion of the buried layer is dominating over the partial recovery (i.e. the compaction) of the upper cap layer. This result is consistent with previous reports on deep implantations [59].

Numerical FEM results, also shown in figure 4, are in good agreement with the experimental measurements for both the as-implanted and annealed sample, with best-fitting values of $\alpha = 4.4 \times 10^{22}$ vacancies cm$^{-3}$ and $D_C = 2 \times 10^{22}$ vacancies cm$^{-3}$, respectively, in satisfactory agreement with the most recent reports [55, 41]. Post-annealing results display a sharper decrease in the swelling at the implantation edge, which is also consistent with the "step-like" variation in material properties at the diamond/graphite interface. The excellent agreement achieved between the experimental and numerical surface swelling estimations provides further confidence in the predictions of internal stress in the implanted diamond.

Shown in figure 5 are the simulation results for the post-annealing case, which is investigated in greater detail using the data from the PL mapping. In figure 5a we report a two-dimensional map of the four computed non-zero stress components (three principal and one shear) in a length/thickness ($xz$) cross-section of the specimen superimposed on the deformed shape of the specimen (where $z$ displacements are multiplied by a factor 2 to highlight the swelling effect). From figure 5a, there is a strong compressive (i.e. positive, according to the sign convention adopted here) stress at the edge of the buried graphitic layer (highlighted in red) due to its constrained expansion. This stress produces a non-negligible effect on the investigated region of the sample near the edge of the implanted region (highlighted in blue), where a tensile stress is observed due to the effect of the nearby swelling.

Figure 5b shows four calculated stress components as a function of distance $d$ from the edge of the implanted region. Consistent with results reported in previous figures, the $x$ direction is perpendicular to the edge of the implanted area, $y$ is parallel to it, and $z$ is the depth direction. The magnitude of the stress is averaged at depths between 0 and 1 μm from the surface, corresponding to our estimate of the probe depth of the PL experiments. As mentioned above, within the probed depth, the main stress component observed is tensile in the $x$ direction which reaches values up to ~10 GPa at distances of 1-2 μm from the implanted area. At these depths, the $y$ stress component is also tensile reaching values of up to ~1.5 GPa, while the $z$ component is compressive due to shear effects and it reaches values of up to ~1 GPa. Also, the shear $xz$ component is not negligible with variations of up to ~2 GPa. Other stress components (such as the remaining shear components, not reported here) are negligible.

Since the Raman spectral features display a significant and well-established dependence on stress for diamond [61], the FEM results derived from the fitting of profilometry measurements were compared to those obtained from the micro-mapping of the first-order diamond Raman line. The shift and broadening of the peak was mapped along a linear scan extending perpendicularly from the edge of the masked side of the implanted area along the $x$ direction (as shown schematically in the inset of figure 6a). From figure 6a, the Raman peak measured in the stressed diamond region adjacent to the implanted region undergoes broadening and a shift determined by the varying stress fields. In particular, the evolution of the peak shift indicates that overall tensile stresses across the sample depth are established between distances 0 μm and ~10 μm, while compressive components in the depth-averaged stresses dominate at distances >10 μm from the edge of the implanted region, until asymptotic zero-shift values are obtained at >30 μm. The dependence of the Raman peak shift $\Delta\omega$ from the hydrostatic stress $\sigma_h$ in diamond is given by the following empirical relation [61]:

$$\Delta\omega = a \cdot \sigma_h + b \cdot \sigma_h^{\,2}$$
, 

(3)

where $a = 2.83$ cm$^{-1}$ GPa$^{-1}$ , $b = -3.65 \cdot 10^{-3}$ cm$^{-1}$ GPa$^{-2}$ . Through rearrangement, we derive the hydrostatic stress relation corresponding to the measured peak shift (as reported in figure 6b) together with the hydrostatic stress predicted through FEM simulations from the combination of the principal stress components ($xx$, $yy$ and $zz$). In contrast to the PL measurements, the Raman data were estimated to originate from a confocal depth of ~3 μm from the sample surface, therefore FEM results were evaluated across this depth and then averaged. It is worth noting that in the Raman measurements, the compressive components arising from deep regions are averaged with tensile components generated from shallow regions thus resulting in a rather complex evolution of the measured "mean stress", as demonstrated by the non-monotonic variation the Raman-derived hydrostatic stress a function of distance from the implanted region. As shown in figure 6b, the agreement between the two datasets is satisfactory from an order-of-magnitude point of view, while the discrepancies between the experimental and numerical estimations of the hydrostatic stress are particularly pronounced where the shear $xz$ stress is not negligible. This is understandable if we consider that (3) is strictly valid only when the principal stresses are applied to the crystal in an exclusively hydrostatic regime, while no empirical coefficients analogous to $a$ and $b$ are reported in literature for the shear stress ($xz$). The observed peak broadening can be ascribed to the non-uniform stress state, which removes the triple degeneracy of the first-order mode resulting in the splitting of the line into three components that could not be spectrally resolved.

Scanning PL mapping of the ZPL NV⁻ emission was performed on the as-implanted sample for the purpose of investigating the properties of the fraction of active luminescent centers that formed during the implantation process (i.e. before the thermal annealing). In particular, regions across the masked and unmasked edges of the implanted area were explored with linear scans in the main growth sector of the crystal as shown in figure 7. In all plots reported in figure 7, the black arrows indicate the scan direction which are consistent with the arrows in the inset pictures of figures 7c, 7d, 7e and 7f. Therefore in all cases, the scans proceed away from the masked and unmasked edges of the implanted area.

Figures 7a and 7b show PL spectra recorded along linear scans at 1 μm steps from the masked and unmasked edges respectively. The spectra are normalized in intensity and displaced along the vertical axis to allow direct observation of the evolution of the peak shape. While the ZPL NV⁻ emission shape remains substantially unchanged in the linear scan along the masked edge, a significant broadening is observed when scanning across the unmasked edge. The broadening increases as the probed point gets closer to the implanted region although a splitting of the $\lambda = 637$ nm emission into different spectral components is not clearly distinguishable (apart from a shoulder at $\lambda = 635.5$ nm). Remarkably, spectral broadening is not observed along the masked edge of the as-implanted sample, which we attribute to the effect of stray implanted ions at the unmasked side of the implanted region. Figures 7c and 7d show the evolution (along a linear scan) of the ratio between the intensity of the ZPL NV⁻ emission and the first order Raman emission for the masked and unmasked edges respectively. The normalization of the PL signal to the first order Raman emission intensity (which should remain constant across the scan) is performed to take into account possible defocusing effects along the scan due to accidental surface tilting. Here we also observe significant differences between scans performed along the masked and unmasked edges. Firstly, the normalized PL intensity is much higher along the unmasked edge with respect to the masked edge while the two values tend to coincide asymptotically at larger distances from the implanted area. Once again we may attribute this observation to a larger concentration of stray ions implanted along the unmasked edge. Secondly, while the normalized PL intensity continues to increase with increasing distance from the implanted region across the masked edge, we observe a non-monotonic trend for the unmasked edge with a pronounced peak at 5-7 μm from the implanted region. This observation may be interpreted using the intensity of PL emission arising from defects in the diamond structure and also follow a non-monotonic trend. At low damage levels, the active PL centers are progressively created while at higher damage densities, the formation of optically active point-defects is increasingly inhibited by the formation of more complex defects and ultimately by the formation of a continuous amorphous network [14]. This effect is visible at the unmasked edge where the spatial variation of the ion induced defect density is broader. Along the masked edge, we only observe the native concentration of active NV⁻ centers which are progressively quenched as they approach the implanted region. As shown in figures 7e and 7f, the trend reported in figure 7c is monotonic on both the masked and unmasked edges as the scan progresses along the low-nitrogen sector highlighted in figure 2. Moreover, in the low-nitrogen "strip" sector we measured similarly low PL intensities from the two edges of the implanted area.

Scanning PL mapping of the ZPL NV⁻ emission was performed in the same experimental conditions on the annealed sample with the purpose of investigating the luminescent centers formed during the thermally-activated process of recombination of nitrogen impurities and ion induced vacancies in close spatial proximity to the heavily stressed implanted region. In this case, the linear scans were performed at 0.5 μm spatial steps.

Figure 8 shows regions of the "main" sector surrounding the implanted region, mapped using 60-μm-long linear scans (#1, 2, 3 and 4). In figure 8, the edges of the implanted region are highlighted (red dashed line) while the coordinates of the starting points of the linear scans with respect to the respective corners of the implanted area are reported in blue. Figures 9a, 9b, 9c and 9d report the normalized PL spectra collected along the four scans reported in figure 8; in each plot, the corresponding scan direction is indicated as a black arrow. As observed in figure 7, the spectra are normalized in intensity and displaced along the vertical axis, to allow a direct observation of the evolution of the peak shape. As expected, after the thermal annealing of the sample, a significant increase of the NV⁻ emission intensity was observed. We attribute this to the formation and/or activation of PL-active centers that are induced straggling ions during the implantation process [53]. Moreover, the splitting of the NV⁻ ZPL emission into up to four different spectral components is more clearly distinguishable with respect to the as-implanted sample. The position of the different components of the split NV⁻ ZPL emission was evaluated by differentiating the spectra reported in figures 9a, 9b, 9c and 9d and identifying the position of the peak maxima in frequency units. The plots of figures 9e, 9f, 9g and 9h report the respective maxima as a function of position. The different spectral components induced by the strong stress fields located around the implanted region are distributed across a spectral range of ~4 THz around the position of the non-perturbed NV⁻ ZPL line (470.9 THz, corresponding to λ = 637 nm). The PL data also follow the local stress tensor and varies spatially due to the implant. The identification of the spectral components is clearer for the scans performed around the masked edge of the irradiated area (scans #1 and #2 reported respectively in figures 9a and 9b) where the transition between implanted and unimplanted regions is sharper. For scans #3 and #4 (figures 9c and 9d), peaks are broader and the identification of spectral components is less straightforward. For this reason, only results obtained from scans #1 and #2 will be commented on further.

Firstly, we observe how the non-perturbed line at 470.9 THz increasingly splits into three main components as the probed spot approaches the corner of the implanted region reaching a position located at 45° with respect to the corner of the implanted area (i.e. the point of intersection of the scans #1 and #2, at distance d = 20 μm with respect to their respective starting points). At this location, the shift in the three components with respect to the non-perturbed line is -0.5 THz, +0.3 THz and +1.0 THz. As the scan progresses from the intersection point and approaches the lateral sides of the implanted area, a fourth component splits from the -0.5 THz peak and progressively evolves into the position of the +0.3 THz component while the -0.5 THz component evolves into a maximum shift of -1.4 THz and subsequently returns to -0.5 THz shift. The +0.3 THz component tends to disappear as the scan progresses from the intersection point until it is replaced by the above-mentioned fourth component. The +1.0 THz component reaches a maximum shift of +2.7 THz, then it stabilizes at +2.3 THz (as measured from scan #1) which yields clearer spectral features as a consequence of the sharper edge of the implanted region.

Similar features, although with reduced detail, can be obtained from scans #3 and #4. As mentioned above, we attribute the lower spectroscopic detail to the fact that the scans are performed along the unmasked edges of the implanted area, where we expect broader and more intense emission.

A clearer identification of the spectral components of the ZPL NV$^-$ emission can be obtained by PL mapping the regions near the implanted area while moving along the "strip" sector highlighted in figure 2. In this sector, the extremely low concentration of nitrogen impurities allows the measurement of much narrower spectral features, as shown in figure 10, where we report a linear scan across the "strip" sector and far away from the implanted area. As with previous plots, the plot in figure 10a shows the spectra normalized in intensity and displaced along the vertical axis to allow a direct observation of the narrowing of the NV$^-$ emission when crossing the low-N sector. As reported in figure 10b, the NV$^-$ emission intensity is also significantly reduced across the low-N sector.

Figure 11 shows the result of PL linear mapping starting at the masked edge of the implanted area, and moving at 1 μm steps away from the implanted area while remaining within the low-nitrogen "edge" sector (as shown schematically in the inset). As for previous images, the black arrows in the plots and in the insets indicate the scan direction. Figure 11a reports normalized PL spectra while figure 11b shows the variation of the spectral positions of the components of the NV$^-$ ZPL line in frequency units as a function of distance from the implanted area. As shown more clearly in a spectrum acquired for a longer integration time at a distance of ~5 μm from the edge of the implanted area (figure 12a), the narrower spectral components of the split NV$^-$ emission are clearly distinguishable and can be suitably fitted with a Lorentzian line function. Table 1 reports the outcomes of the fitting procedure.

**Table 1.** Center frequency, linewidth and relative intensity of the Lorentzian fitting curves (a1-a5, b1) reported in figure 12a and 12b, respectively; the intensity values are expressed as percentages of the overall intensity, evaluated by integrating the emission spectrum after background subtraction. The formula f(x)=2A/pi*w/[4*(x-xc)^2+w^2] was employed for the Lorentzian peak fitting.

|  | Center (THz) | Width (GHz) | Intensity (%) |
|---|---|---|---|
| Peak a1 | $468.108 \pm 0.006$ | $(80 \pm 2) \cdot 10^1$ | 15 |
| Peak a2 | $469.886 \pm 0.006$ | $(94 \pm 2) \cdot 10^1$ | 22 |
| Peak a3 | $470.947 \pm 0.002$ | $545 \pm 8$ | 28 |
| Peak a4 | $472.012 \pm 0.004$ | $(65 \pm 2) \cdot 10^1$ | 23 |
| Peak a5 | $472.87 \pm 0.02$ | $(106 \pm 5) \cdot 10^1$ | 12 |
| Peak b1 | $471.0 \pm 0.1$ | $126 \pm 2$ | 100 |

The splitting into up to five spectral components monotonically increases as the distance from the heavily stressed implanted area decreases, starting at about 15 μm from the edge and covering a range of ~4.8 THz i.e. a broader splitting range with respect to what observed in the "main" sector (see figure 9 for comparison). As a comparison, in figure 12b we report the PL spectrum acquired from a region at a distance of >20 μ, which we can assume to be unaffected by the stress field. As expected, a sharp single emission line is observed whose Lorentzian fitting parameters are also reported in table 1.

G. Davies *et al.* derived from absorption measurements under uniaxial compression the values of the linear coefficients linking the position of the split components of the NV⁻ ZPL (|EX> and |EY> states) and the applied stress [12]. If we apply such coefficients to the values of the different components of the complex stress field obtained from the numerical modeling (see section 3), it is possible to derive trends of the energy shifts for the different spectral components of NV⁻ emission versus the distance from the implanted region. Table 2 reports the values of the coefficients, after conversion from energy to frequency units.

**Table 2.** linear coefficients (expressed in kHz·Pa⁻¹) linking the frequency shift of |EX> and |EY> states (labeling the columns) to applied stress in the <100>, <110> and <111> directions (labeling the rows), as derived from optical absorption measurements of the ZPL absorption of the NV⁻ center in σ-polarization [12]; the coefficients highlighted in bold characters have been successfully employed to interpolate the experimental data.

|       | \|EX>   | \|EY>   |
|-------|---------|---------|
| <100> | **+0.919** | **-0.109** |
| <110> | +0.648  | **-0.766** |
| <111> | -       | +0.389  |

The resulting trends, as reported in the continuous line plots in figure 11b, are satisfactorily compatible with experimental data, particularly considering that the numerical model of the stress field has been derived under significant assumptions. These include: (i) the same functional dependence of all mechanical properties with vacancy density, based on the outlined phenomenological model and (ii) a simplified geometry of the buried graphite layer with respect to the real experimental situation. Moreover, the deconvolution of the multiple spectral components arising from stress fields along different <100>, <110> and <111> directions is not as unequivocal as in uniaxial stress experiments. Despite these assumptions, we observe that the stress <100> components in the $x$ and $y$ directions (i.e. where the stress field is more intense) can account for the observed trends. In the case of the |EX> component arising from stress in the <100> direction (red curve in figure 11b), we have no experimental data at small distances from the implanted area where the numerical results would predict much larger shifts. Finally, we tentatively attribute the large shift at low frequency to a shift of the |EY> state induced by a stress field in the <101> direction. No unequivocal attribution could be found to relate the measured shifts to stresses in the <111> direction. Given the complexity of the observed effect, a fully detailed quantitative analysis of the observed trend is beyond the scopes of the present work and will be carried out in future systematic investigations.

## 5. Conclusions

We report the systematic investigation of the splitting of NV⁻ photoluminescent centers induced in diamond by the stress field caused by MeV ion implantation. Engineering the strain field in implanted diamonds opens new opportunities for NV⁻ ensemble-based devices, by spectrally separating the inequivalent orientations and providing permanent shifts in the spectral features affording some degree of tuning.  Ion-beam induced stress fields are caused by the formation of the stable amorphous or graphitic phase forming in the highly-damaged, sub-superficial region, respectively before and after thermal annealing. The ion-induced density variation creates complex internal stress fields in the unimplanted surrounding regions, resulting in the splitting of the NV⁻ emission into different spectral components which have a range of ~4.8 THz. The employment of implantation masks proved to be beneficial to improve the sharpness of the edges of the implanted areas. This allowed the definition of well-defined boundary geometries and in addition to the PL mapping of the NV⁻ spectral features with micrometric spatial resolution in low-nitrogen sectors in the diamond crystal.

Consistent with modeling in this work, the stress fields occurring in the implanted and annealed sample were simulated with FEM numerical codes by adopting a simple semi-analytical model developed in previous works [54, 47]. The free parameters in the model were determined by optimizing the consistency with surface swelling data while the model was further validated by testing its consistency with experimental Raman measurements.

The measured stress distributions at the edges of the implanted region were coupled with known coefficients linking static stress and NV⁻ sub-level shifts [12], along with the resulting values of the predicted line splitting proved to be in satisfactory agreement with the experimental data. A fully unequivocal attribution could not be established at this stage, and it will be the subject of future investigations.

**Acknowledgments**

The rendered image in the inset of figure 4 is courtesy of A. Cirino. The work of P. Olivero is supported by the "FIRB - Futuro in Ricerca 2010" project (CUP code: D11J11000450001), which is gratefully acknowledged. A. D. G. is supported by an Australian Research Council Queen Elizabeth II Fellowship (Project number DP0880466). B. C. G. is supported by an Australian Research Council Future Fellowship (Project number FT110100225).

**Figures**

Figure 1

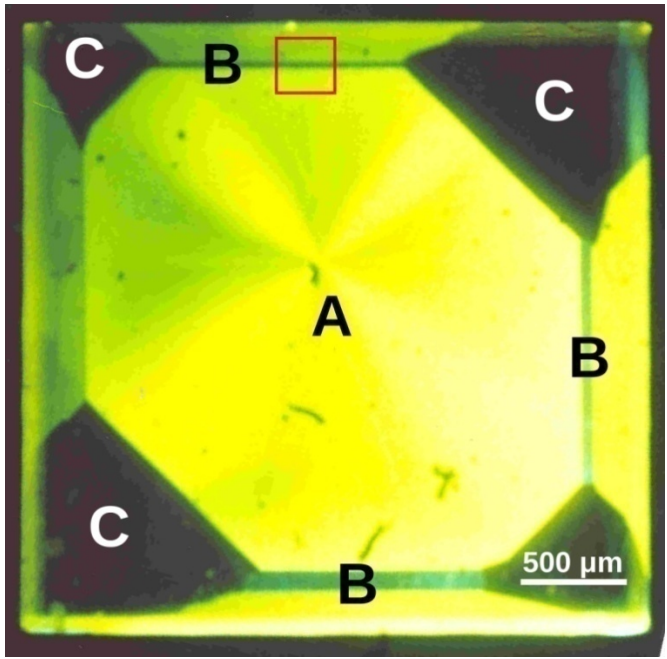

Figure 2

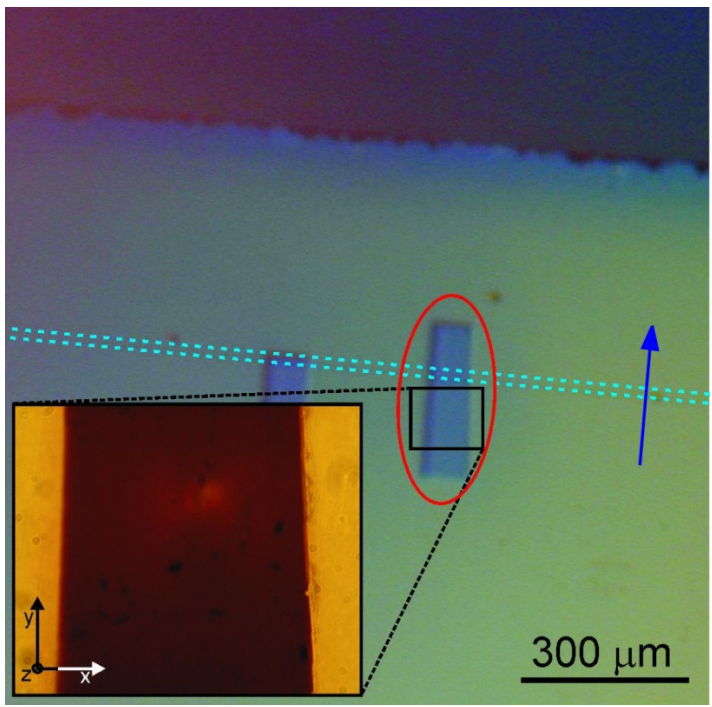

Figure 3

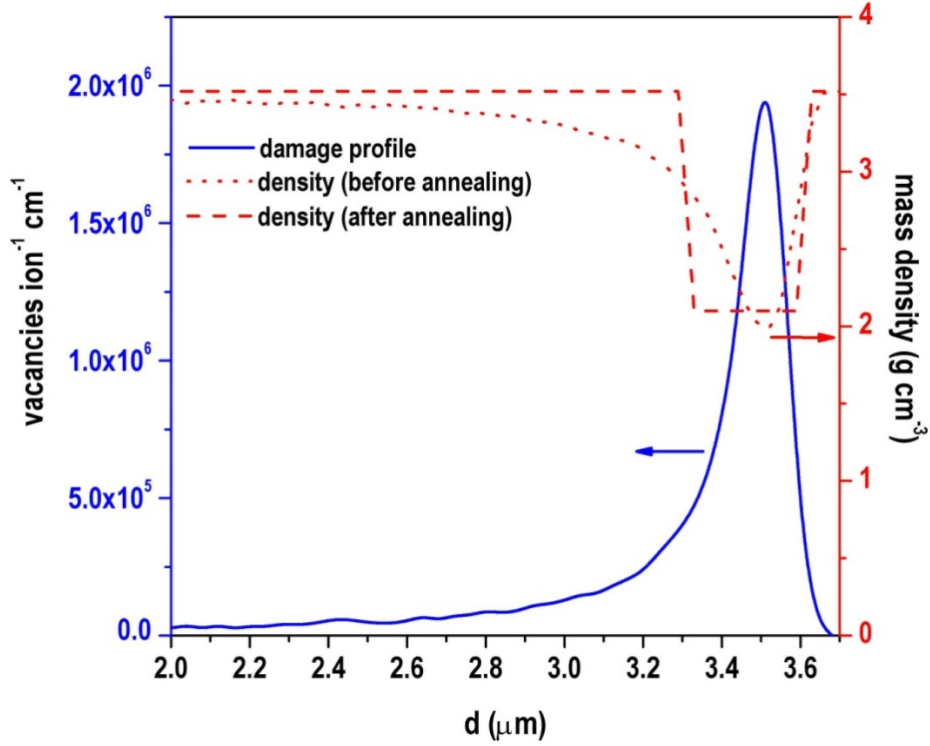

Figure 4

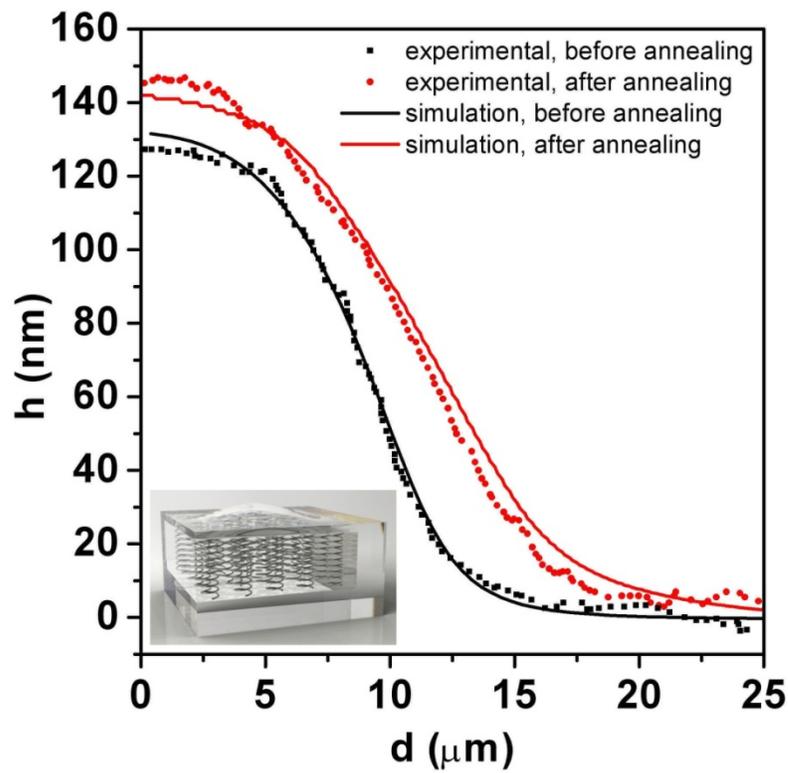

Figure 5

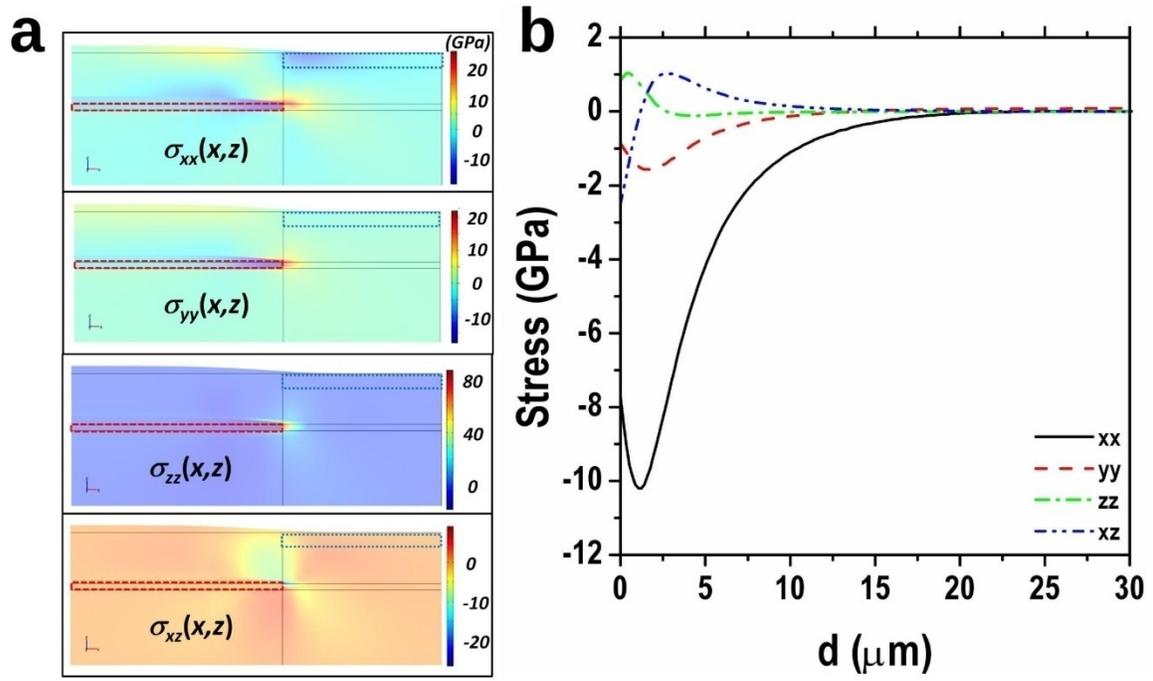

Figure 6

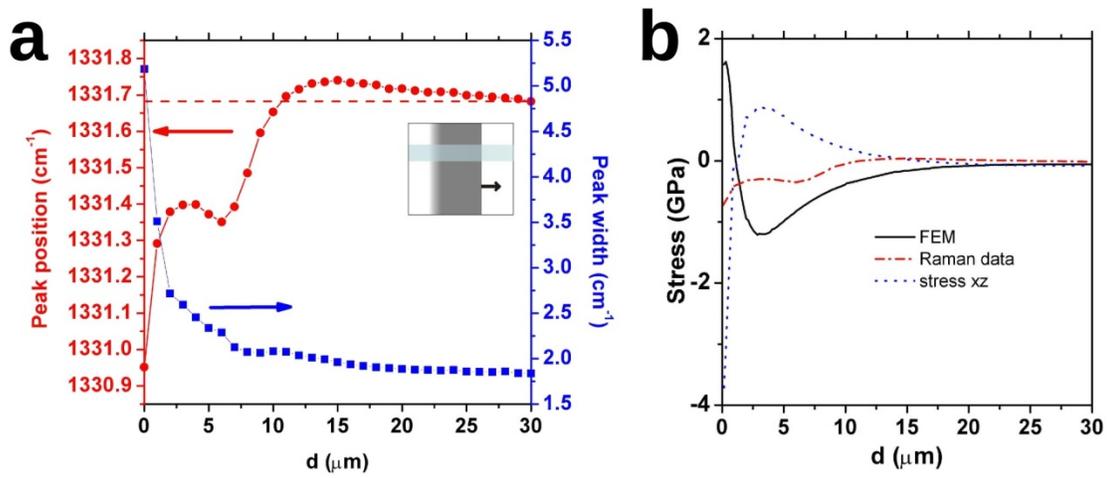

Figure 7

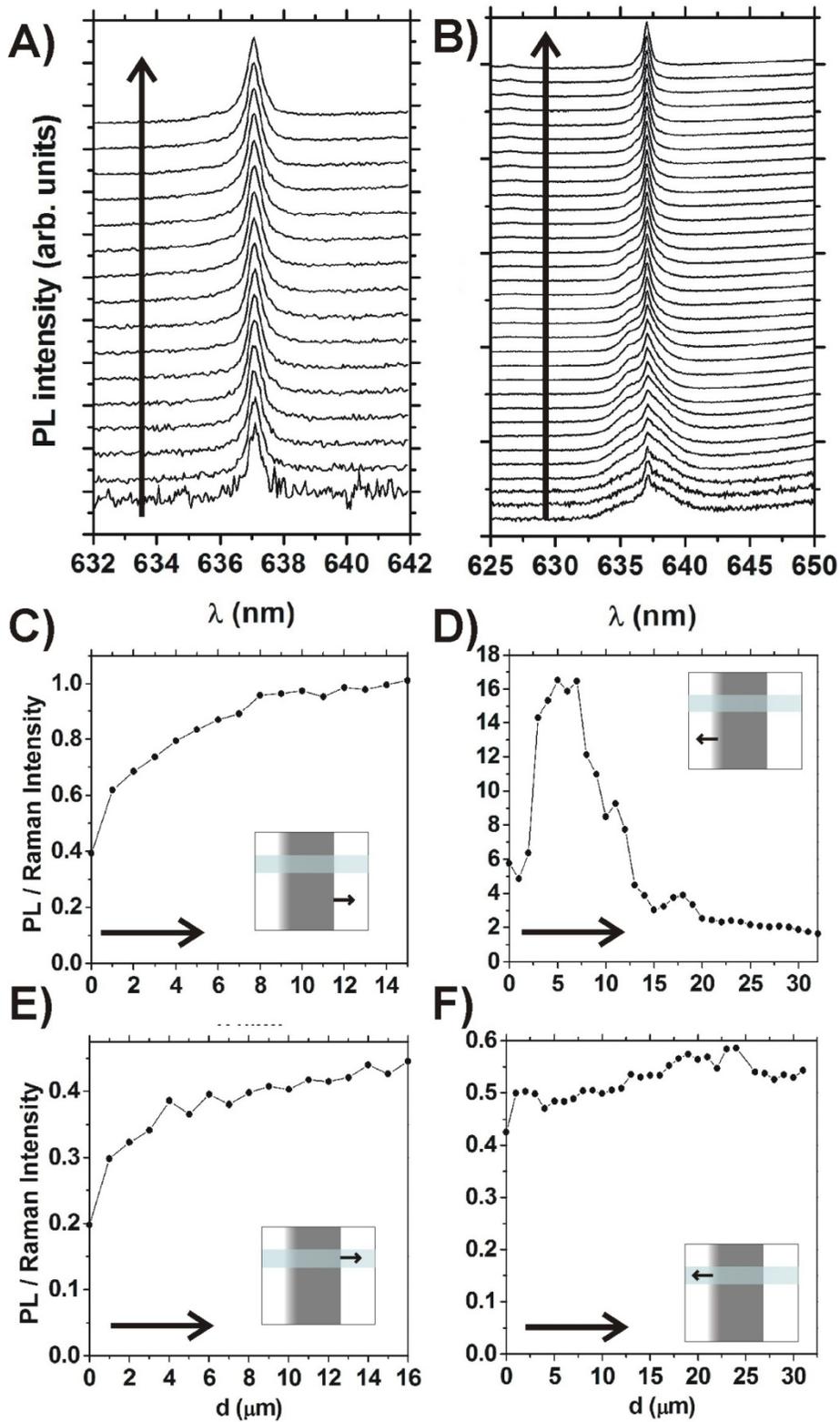

Figure 8

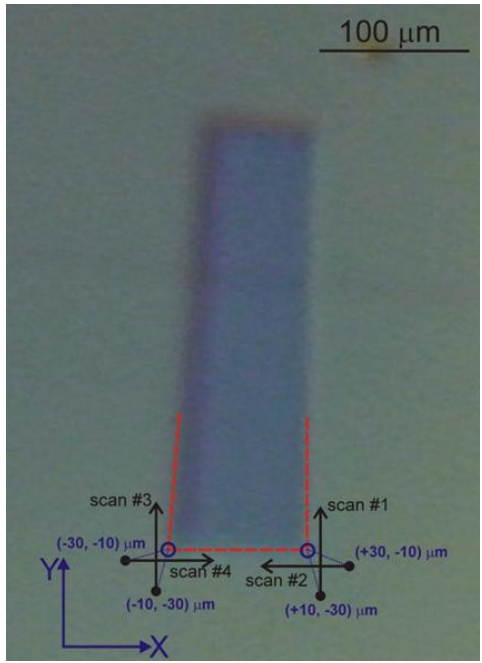

Figure 9

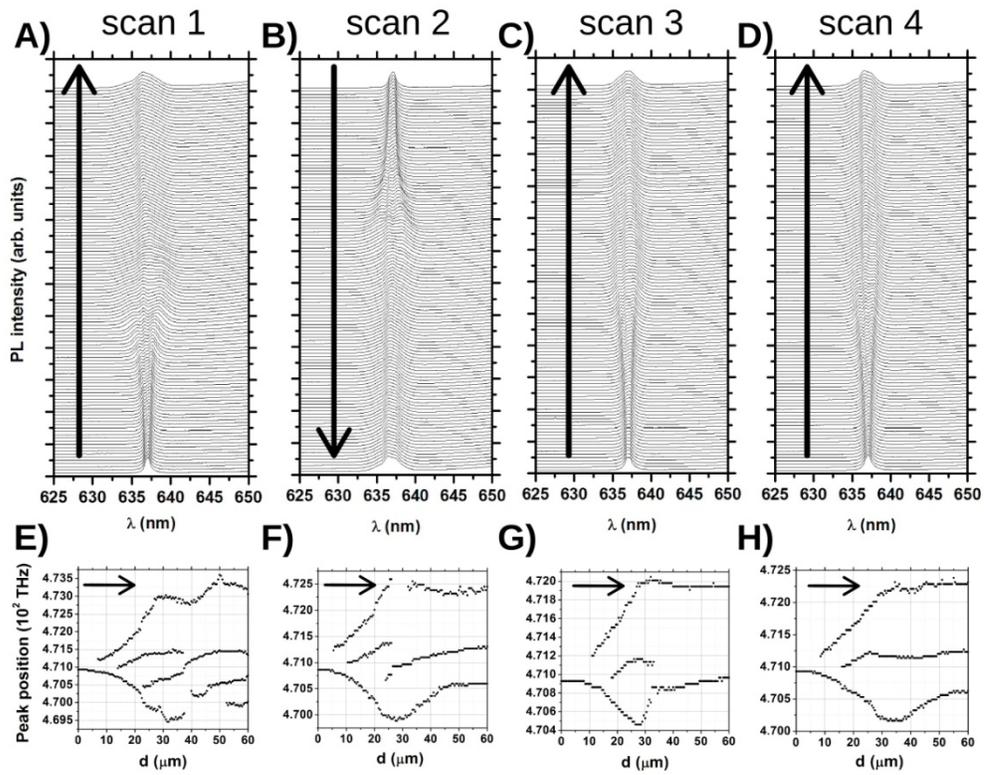

Figure 10

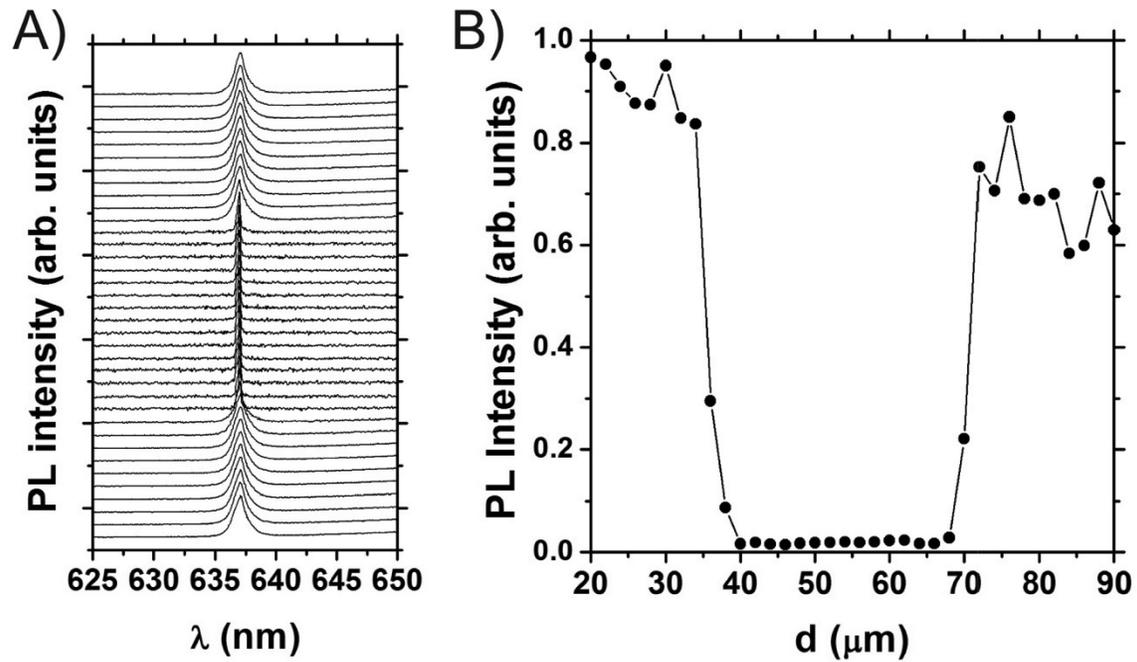

Figure 11

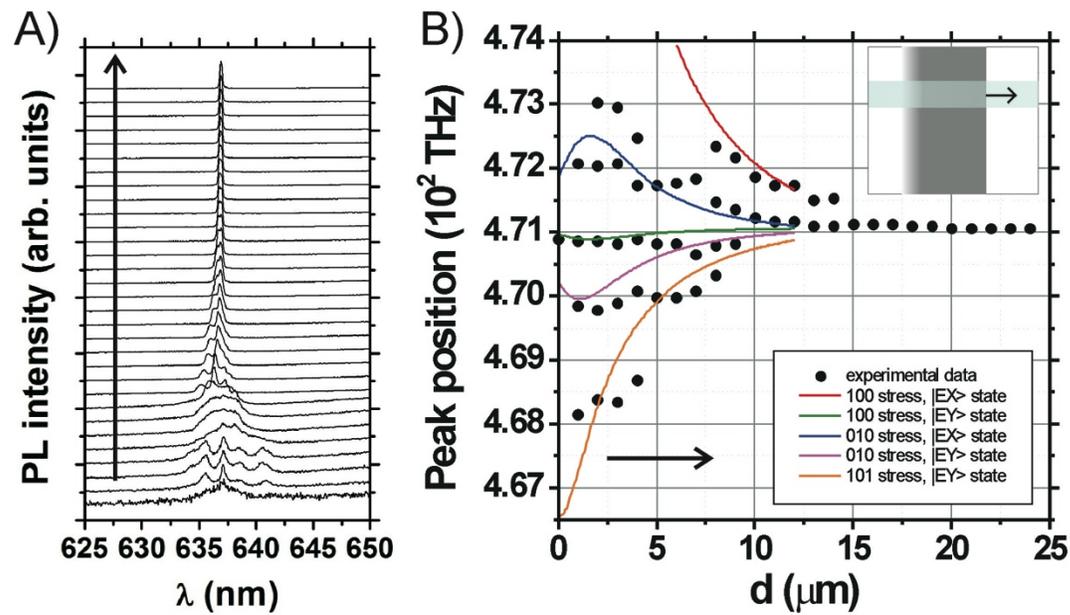

Figure 12

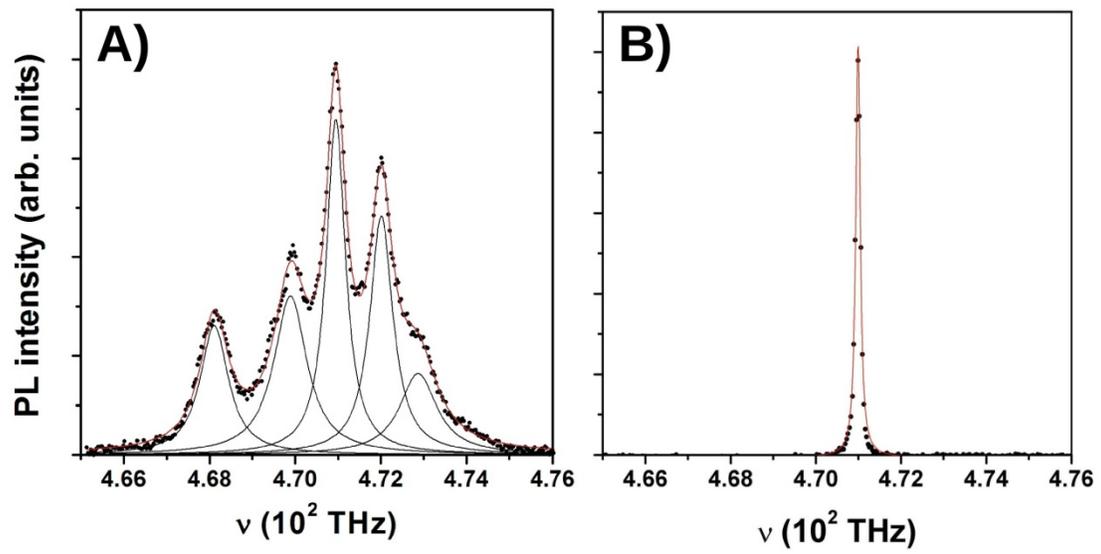

**Figure captions**

Figure 1 (color online): Cathodoluminescence map of the type Ib HPHT diamond crystal used for these experiments; sectors are clearly visible since different sectors are characterized by different impurity concentrations. By comparison with the "main" sector (A), the "strip" sectors extending near the sample edges (B) are characterized by lower nitrogen concentration, while the sectors near the sample corners (C) contain a higher concentrations of impurities. The red square highlights the area of interest, across the top "strip" sector.

Figure 2 (color online): Optical transmission microscopy image of the implanted region (highlighted in the red oval) extending across the "main" sector (bottom) and the low-nitrogen "strip" (highlighted in blue); the inset shows a zoom image of the implanted area, where it can be appreciated that the masked side of the implantation (right side) is sharper than the unmasked side (left side). The blue arrow highlights the direction of the PL scan to locate the "strip" sector reported (see figure 10).

Figure 3 (color online): SRIM Monte Carlo simulation of the damage profile induced in diamond by 2 MeV He$^+$ ions at a fluence of $1\times10^{17}$ cm$^{-2}$ (blue continuous line, left axis): the damage is mostly concentrated at the end of the range, namely ~3.5 µm below the sample surface; variation of mass density derived from the damage profile, before and after thermal annealing (red dotted and dashed lines respectively, right axis).

Figure 4 (color online): Comparison between the experimentally measured and numerically-calculated (FEM) surface. The surface swelling values at the edge of the implanted area, both before and after annealing; the zero position corresponds to the edge of the implanted area; inset: schematic representation of the basic mechanism underlying the swelling process.

Figure 5 (color online): a) Spatial distribution of numerically-calculated principal ($xx$, $yy$, $zz$) and shear ($xz$) stresses (in GPa) in the considered specimen after annealing; the red dashed rectangle indicates the buried graphitic layer while the blue dotted rectangle the region where stresses are evaluated and correlated to PL measurements; b) Plots of the stress components at the edge of the implanted area; consistent with the estimated probing depth of PL measurements, data are evaluated between 0 and 1 µm and then averaged.

Figure 6 (color online): a) Mapping of the first-order Raman peak shift (red circles, left axis) and broadening (blue squares, right axis) as collected from a linear scan extending perpendicularly from the masked edge of the implanted region, as schematically shown in the inset; the position of the non-stressed line is highlighted by the red dashed line; b) Hydrostatic stress derived from the Raman peak shift data (red dotted curve) and from principal stress components estimated from the FEM simulations (continuous black line). Consistent with the estimated probing depth of Raman measurements, the FEM data are evaluated between 0 and 3 μm and then averaged. We attribute the discrepancy to the presence of a shear stress component ($xz$), also included in the plot (blue dotted curve).

Figure 7 (color online): Results of PL scanning across the masked (A, C, E) and unmasked (B, D, F) edges of the as-implanted region. In all plots, the black arrows indicate the scan direction, consistent with the arrows in the inset images. A), B): normalized PL spectra collected at 1 μm steps away from the masked and unmasked edges of the implanted area respectively, as shown in the insets of figures 5c and 5d respectively; the spectra are displaced along the vertical axis for readability; C), D): evolution of the ratio between the intensity of the ZPL NV$^-$ emission and the first order Raman emission along the linear scans in the "main" sector are schematized in the insets; E), F): evolution of the NV$^-$ : Raman ratio along the linear scans in the "strip" sector schematized in the insets.

Figure 8 (color online): Optical microscopy image of the implanted area: the arrows indicate the length and direction of linear scans #1, #2, #3 and #4; the edges of the implanted area are highlighted in red dashed line, while the coordinates of the starting points of the linear scans with respect to the respective corners of the implanted area are reported in blue.

Figure 9: Results of PL scanning around the implanted area of the annealed sample, as reported in figure 8. In each plot, the black arrows indicate the scan direction, consistent with that reported in figure 8. A), B), C), D): normalized PL spectra collected at 0.5 μm steps along the scans #1, 2, 3 and 4, respectively, as reported in figure 8; the spectra are displaced along the vertical axis for readability; E), F), G), H): evolution of the split components of the NV$^-$ ZPL line as a function of position along scans #1, #2, #3 and #4, respectively.

Figure 10: A) normalized PL spectra collected at 2 μm steps away from the implanted region and across the low-nitrogen "edge" sector (see the blue arrow in figure 2); the spectra are displaced along the vertical axis for readability; B) intensity of the ZPL $NV^-$ emission along the linear scan; the extension of the "edge" sector can be clearly distinguished by the significant sharpening (A) and the decrease in intensity (B) of the $NV^-$ emission.

Figure 11 (color online): A) normalized PL spectra collected by moving across a linear scan at 1 μm steps from the edge of the implanted region within the low-nitrogen "edge" sector, as shown schematically in the inset; the spectra are displaced along the vertical axis for readability; B) (dots) evolution of the split components of the $NV^-$ ZPL line as a function of the distance from the edge of the implanted area; (lines) trends predicted by the numerical model of the internal stress fields integrated with the coefficients reported in [12]. In all plots, the black arrows indicate the scan direction, consistent with that reported in the inset of B).

Figure 12: A) PL spectrum (dots) acquired at a distance of ~5 μm from the edge of the implanted area; up to five spectral components are distinguishable, as highlighted by the fitting of multiple Lorentzian functions (lines). B) PL spectrum (dots) acquired at a distance of >20 μm from the edge of the implanted area; a single sharp spectral component is observed, as highlighted by the fitting Lorentzian curve (line).